\documentclass[twocolumn,aps,prl,amsmath,amssymb]{revtex4-1}
\usepackage{graphicx}% Include figure files
\usepackage{dcolumn}% Align table columns on decimal point
\usepackage{bm}% bold math
\usepackage{epstopdf}
\DeclareGraphicsExtensions{.pdf,.eps,.png,.jpg,.mps}

\begin{document}

\preprint{AIP/123-QED}

\title{Photo-thermoelectric and photoelectric contributions to light detection in metal-graphene-metal photodetectors}
\author{T.J. Echtermeyer$^1$}
\author{P.S. Nene$^1$}
\author{M. Trushin$^{2,3}$}
\author{R.V. Gorbachev$^4$}
\author{A.L. Eiden$^1$}
\author{S. Milana$^1$}
\author{Z. Sun$^1$}
\author{J. Schliemann$^3$}
\author{E. Lidorikis$^5$}
\author{K.S. Novoselov$^4$}
\author{A. C. Ferrari$^1$}

\email{acf26@eng.cam.ac.uk}

\affiliation{$^1$Cambridge Graphene Centre, University of Cambridge, Cambridge CB3 0FA, UK}
\affiliation{$^2$Department of Physics, University of Konstanz, D-78457 Konstanz, Germany}
\affiliation{$^3$Institute for Theoretical Physics, University of Regensburg, D-93040 Regensburg, Germany}
\affiliation{$^4$School of Physics $\&$ Astronomy, University of Manchester, Oxford Road, Manchester M13 9PL, UK}
\affiliation{$^5$Department of Materials Science and Engineering, University of Ioannina, Ioannina, Greece}

\begin{abstract}
Graphene's high mobility and Fermi velocity, combined with its constant light absorption in the visible to far-infrared range, make it an ideal material to fabricate high-speed and ultra-broadband photodetectors. However, the precise mechanism of photodetection is still debated. Here, we report wavelength and polarization dependent measurements of metal-graphene-metal photodetectors. This allows us to quantify and control the relative contributions of both photo-thermo- and photoelectric effects, both contributing to the overall photoresponse. This paves the way for a more efficient photodetector design for ultra-fast operating speeds.
\end{abstract}

\maketitle
The unique optical and electronic properties of graphene make it ideal for photonics and optoelectronics\cite{Bonaccorso2010}. A variety of prototype devices have already been demonstrated, such as transparent electrodes in displays\cite{Bae2010} and photovoltaic modules\cite{dearco2010}, optical modulators\cite{Liu2011}, plasmonic devices\cite{Ju2011,Echtermeyer2011,Liu2011,Schedin2010,Fei2012,Chen2012}, microcavities\cite{Engel2011,Furchi2012} and ultra-fast lasers\cite{Sun2010}. Amongst these, a significant effort has been devoted to photodetectors\cite{Lee2008,Mueller2009,Park2009,Mueller2010,Urich2011,Xia2009,Lemme2011,Gabor2011,Xu2010,Konstantatos2012,Furchi2012,Engel2011,Echtermeyer2011,Freitag2013,Yan2013,Pospischil2013}.

Various photodetection schemes and architectures have been proposed to date. The simplest configuration is the metal-graphene-metal (MGM) photodetector, in which graphene is contacted with metal electrodes as source and drain\cite{Lee2008,Mueller2009,Park2009,Mueller2010,Urich2011,Xia2009}. These detectors can be combined with metal nanostructures enabling local surface plasmons and increased absorption, realizing an enhancement in responsivity (i.e. the ratio of the light-generated electrical current to the incident light power)\cite{Echtermeyer2011,Liu2011_2}. Microcavity based photodetectors were also used, with increased light absorption at the cavity resonance frequency, again achieving an increase in responsivity\cite{Engel2011,Furchi2012}. Another scheme is the integration of graphene with a waveguide to increase the effective interaction length with light\cite{Kim2011,Pospischil2013}. Hybrid approaches employ semiconducting nanodots as light absorbing media\cite{Konstantatos2012}. In this case, light excites electron-hole (e-h) pairs in the nanodots. The electrons are trapped in the nanodot, while the holes are transferred to graphene, thus effectively gating it\cite{Konstantatos2012}. Under applied drain-source bias this results in a shift in the Dirac-point, thus a modulation of the drain-source current\cite{Konstantatos2012}. Due to the long trapping time of the electrons within the dot, the transferred holes can cycle many times through the phototransistor before relaxation and e-h recombination. This gives a photoconductive-gain, i.e. one absorbed photon effectively results in an electrical current of several electrons. Responsivities$>$10$^7$ A/W were reported\cite{Konstantatos2012}, but with a ms timescale, not suitable for e.g. high-speed optical communications. Devices were also fabricated for detection of THz light\cite{Vicarelli2012,THz2}. In this low energy range, Pauli-blocking forbids the direct excitation of e-h pairs due to finite doping. Instead, an antenna coupled to source and gate of the device excites plasma waves within the channel. These are rectified, leading to a detectable dc output voltage\cite{Vicarelli2012,THz2}. Photodetectors based on intrinsic graphene plasmons were also demonstrated\cite{Freitag2013}. Graphene, structured into periodic nanoribbons (GNRs), forms a plasmonic metamaterial enabling standing plasmons excitation by infrared light. These lead to an increase of the electron and phonon temperatures, which causes a detectable change of the electrical conductivity of graphene\cite{Freitag2013}.

\begin{figure}
\centerline{\includegraphics[width=90mm]{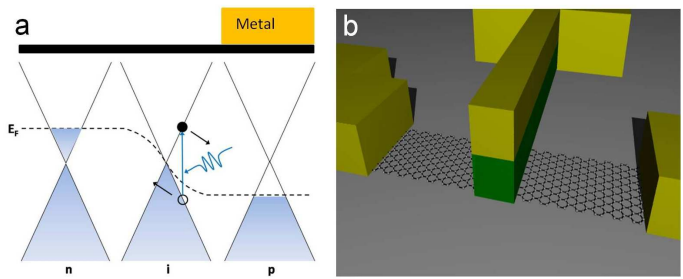}}
\caption{Overview of a) energy-band profile in MGM photodetectors and b) transistor-like graphene-based photodetector employing a top gate.}
\label{graph:fig1pd}
\end{figure}

MGM photodetectors play an important role because they are easy to fabricate, not relying on nanoscale lithography. They operate over a broad wavelength range as the light-matter interaction is mostly determined by graphene itself. Further, ultrahigh operating speeds can be achieved\cite{Mueller2010}, as no bandwidth limiting materials are employed\cite{Konstantatos2012}. MGM photodetectors can be considered as the fundamental building block for the other architectures mentioned above. They consist of a graphene channel contacted by two electrodes, either of the same\cite{Lee2008,Mueller2009,Park2009,Xu2010,Urich2011,Xia2009} or two different metals\cite{Mueller2010}. The difference in workfunction between the metal pads and graphene leads to charge transfer\cite{Giovanetti2008}, with a consequent shift of the graphene Fermi level in the region below the metal pads\cite{Giovanetti2008}. The Fermi level gradually moves back to that of the un-contacted graphene when crossing from the metal covered region to the metal-free channel\cite{Mueller2009}. This results in a potential gradient extending$\sim$100-200nm from the end of the metal pad to the metal-free graphene channel\cite{Mueller2009}. This inhomogenous doping profile creates a junction along the graphene channel. This can in principle be a pn-, nn- or pp-junction between the graphene underneath the channel and within the channel, as the channel Fermi-level can be controlled by the back gate. Fig.\ref{graph:fig1pd}a shows a schematic of the doping profile induced by the metal contact. The formation of this junction is crucially important in the photodetection process, as it results in an internal electric field, capable of separating the light induced e-h pairs.

Another approach to create such junctions, is to exploit a dual-gate transistor structure (Fig.\ref{graph:fig1pd}b)\cite{Lemme,Lemme2011,Gabor2011}. The simultaneous electrostatic doping of the graphene channel by means of a global bottom- and a local top-gate allows formation of nnn, ppp, npn, or pnp junctions, respectively. From an application point of view, the dual gate structure requires more fabrication steps, as well as more supply voltages for the gating, but allows control of the doping levels on both sides of the junction. The MGM photodetector requires fewer processing steps, but has a fixed doping level underneath the contacts, thus allowing fewer operational degrees of freedom. However, it is more suited for applications, due to the simpler fabrication; the single back gate can be used to control the potential gradient in the vicinity of the contacts.

Currently, two effects are thought to contribute to the photoresponse in graphene-based photodetectors, both requiring spatially in-homogenous doping profiles: photo-thermoelectric\cite{Lemme2011,Gabor2011,Song2011,Park2009,Freitag2013} and photoelectric\cite{Mai2011,Lee2008,Mueller2009,Mueller2010,Xia2009}. The photo-thermoelectric effect results from local heating of, e.g., the pn-junction, due to the incident laser power. Non-equilibrium hot carriers are excited with an electron temperature (T) higher than that of the lattice\cite{Song2011}. Different doping levels on both sides of the junction give different Seebeck coefficients\cite{Song2011}. These are a measure of the induced thermoelectric voltage due to a T gradient, and depend on the Fermi-level according to Mott's formula\cite{Xu2010,Gabor2011,Song2011,Ashcroft1976}:
\begin{equation}
S = -\frac{\pi^2 k_B^2 T}{3q} \frac{1}{\sigma} \frac{d\sigma}{d\mu}
\label{eq2}	
\end{equation}

with $k_B$ the Boltzmann constant, $\sigma$ the conductivity, $q$ electron charge, and $\mu$ the chemical potential. As a consequence, a net electron flow results\cite{Xu2010,Gabor2011,Song2011,Ashcroft1976}, producing a photo-thermoelectric voltage $V_{PTE}$ \cite{Xu2010,Gabor2011,Song2011,Ashcroft1976}:
\begin{equation}
V_{PTE} = \left(S_1-S_2\right) \ \Delta T
\label{eq1}	
\end{equation}
with $\Delta T$ the T increase of the hot electrons within the junction, and $S_1,S_2$ the Seebeck coefficients of the two regions with different doping at the junction.

Due to the non-monotonous dependence of the difference of the Seebeck coefficients in the two differently doped regions of the junction, the resulting $V_{PTE}$ exhibits multiple sign reversals in dependence of the gate voltage\cite{Gabor2011,Song2011}. This results in a six-fold pattern, due to the non-monotonic S in a plot of the photovoltage in dependence of the two doping levels on either side of the junction, as theoretically proposed in Ref.\cite{Song2011} and experimentally observed in Ref.\cite{Gabor2011}.

Besides photo-thermoelectric effects, light induced heating of one contact can also lead to a T gradient, resulting in a photo-thermoelectric contribution to the photovoltage, as that described in Refs.\cite{Zuev2009,Wei2009}, where a T gradient was created employing a microfabricated heater\cite{Zuev2009,Wei2009}.

The presence of the junction in the photo-thermoelectric effect is as important as in the photoelectric effect. The potential gradient within the junction separates the light induced e-h pairs and leads to a current flow as in a conventional photodiode\cite{Mai2011,Sze1981}. However, to the best of our knowledge, direct evidence and quantification of the photoelectric effect contribution to the photovoltage generation is still missing.

Here we investigate the wavelength and polarization dependent responsivity of MGM photodetectors. The measured light polarization dependent responsivity, combined with the spatial origin of the photoresponse obtained from photovoltage maps, allows us to determine the photoresponse mechanisms and quantitatively attribute it to photo-thermo- and photo-electric effects.

\begin{figure}
\centerline{\includegraphics[width=75mm]{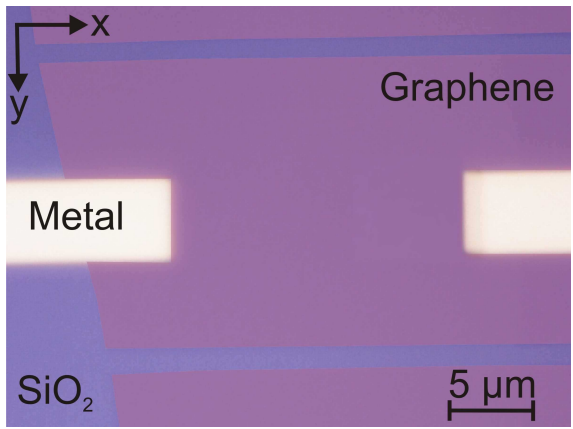}}
\caption{Optical micrograph of device. Graphene is contacted with two metal electrodes.}
\label{graph:fig1}
\end{figure}

Our devices are fabricated as follows. Graphene is produced by mechanical exfoliation of graphite (NGS Naturgraphit GmbH) on top of Si+SiO$_2$ (300nm)\cite{Novoselov2004,Novoselov2005} and its single layer nature confirmed by optical microscopy\cite{Casiraghi2007} and Raman spectroscopy\cite{Ferrari2006,Ferrari2013}. E-beam lithography is used to define the contacts, followed by e-beam evaporation of the contact metal, consisting of a 4nm Ti adhesion layer, and 80nm gold Au pads, using lift-off to ensure good mechanical adhesion as well as good electrical contact. Fig.\ref{graph:fig1} shows an optical micrograph of a representative device. The two metal contacts with a width of 5$\mu m$ face each other. The highly-doped ($\rho=0.001-0.005\ \Omega cm$) Si back gate allows us to control the Fermi-level in the graphene channel.

Photovoltage mapping is performed at 457, 488, 514, 633, 785, and 1550nm (laser power P$<$1mW). The samples are bonded into a chip-carrier, and connected in a two-terminal configuration to a Keithley Nanovoltmeter 2182A with an additional sourcemeter controlling the gate voltage. The position dependent generated photovoltage is monitored while linearly polarized laser light with diffraction limited spot size is scanned over the device. Light from the laser sources is focused through a 100x ultra-long working distance objective (NA=0.6) on to the photodetectors. A Fresnel-Rhomb polarizer allows us to rotate the light polarization. Polarization control at 1550nm is achieved employing a $\frac{\lambda}{2}$ wavelength plate. Raman measurements are carried out using a Renishaw inVia spectrometer with P$<$1mW to avoid any possible damage. This allows monitoring defects\cite{Cancado2011,Ferrari2000,Ferrari2007,Ferrari2013}, as well as local doping\cite{Pisana2007,Das2008,Echtermeyer2011,Ferrari2013}.
\begin{figure}
\centerline{\includegraphics[width=90mm]{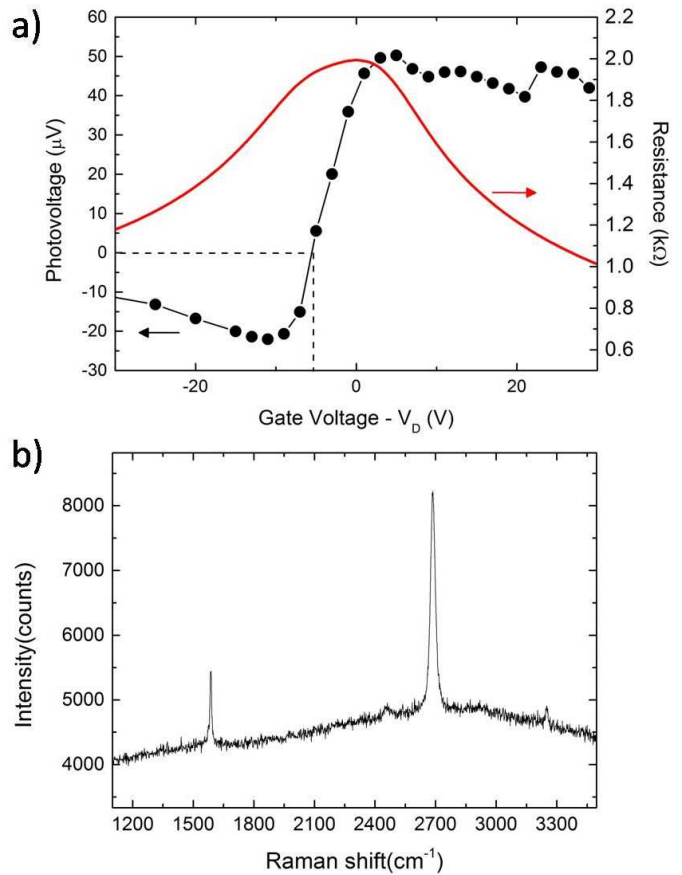}}
\caption{a) Photovoltage and resistance as a function of back gate voltage. b) Raman spectrum measured at 514.5nm and $V_g - V_D$=-5V, corresponding to the voltage at which the photovoltage exhibits a sign reversal in a).}
\label{graph:figdoping}
\end{figure}

The doping of the pn-junction can be determined by measuring the back gate voltage dependence of the photoresponse. Fig.\ref{graph:figdoping}a compares the photovoltage in dependence of back gate voltage $V_{g}$ with the resistance, at an incident light wavelength of 633nm. The photovoltage shows a sign reversal at -5V relative to the Dirac point, $V_D$. The photovoltage is zero at $V_g - V_D$=-5V, as the doping underneath the metal contact and in the non-contacted graphene is equal, meaning that no junction is present, thus no photovoltage can be produced. The point of vanishing photovoltage lies in the p-doped branch of the resistance curve. From the carrier density $n = \epsilon \epsilon_0 \frac{V_{g}-V_D}{e \ t}$, with t the oxide thickness, the Fermi-level $E_F = \hbar v_F \sqrt{\pi n}$ can be derived to be$\sim$60meV\cite{Novoselov2004,Novoselov2005}. This p-doping underneath the metal contact is independent of back gate voltage, due to Fermi-level pinning\cite{Giovanetti2008}. For photovoltage mapping, the back gate voltage is set to $V_g - V_D$=+10V to achieve n-doping$\sim$100meV in the non-contacted graphene, thus a pn-junction. The doping values are in good agreement with what can be estimated from the Raman spectrum in Fig.\ref{graph:figdoping}b, measured at $V_g - V_D$=-5V. The spectrum shows no D peak, indicating negligible defects\cite{Ferrari2006,Ferrari2013,Cancado2011,Ferrari2000}. The 2D peak is a single sharp Lorentzian with full width at half maximum, FWHM(2D)$\sim$28cm$^{-1}$, signature of single layer graphene (SLG). The G peak position, Pos(G), and full width at half maximum, FWHM(G), are$\sim$1587cm$^{-1}$ and$\sim$9cm$^{-1}$. Pos(2D)$\sim$2686cm$^{-1}$, and the 2D to G intensity and area ratios, I(2D)/I(G) and A(2D)/A(G), are 3.1 and 8.8 respectively. This indicates p-doping$<$100meV\cite{Das2008}, confirming the electrical characterization. Further, while pristine SLG absorbs 2.3\% of the incident light\cite{Nair2008}, doping can significantly decrease the absorption by Pauli blocking\cite{Mak2008,Lagatsky2013}. However, the estimated low doping level$\sim$100meV derived from the electrical and Raman measurements does not induce any absorption decrease in the wavelength range used in this work.
\begin{figure}
\centerline{\includegraphics[width=80mm]{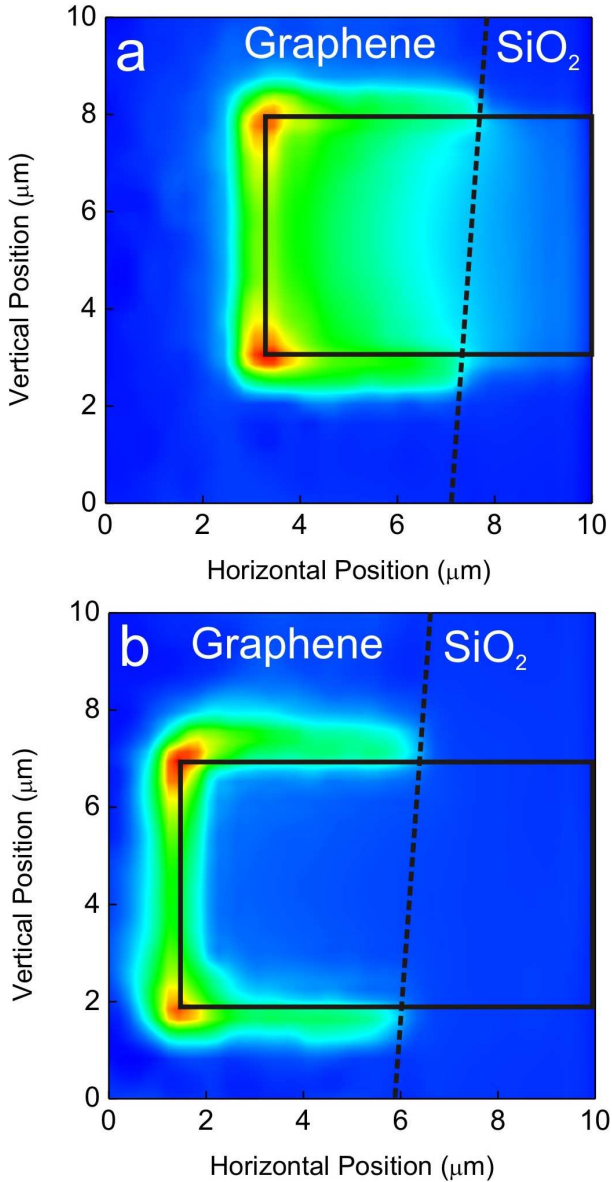}}
\caption{Photovoltage maps for a) 457nm and b) 785nm excitations}
\label{graph:fig2}
\end{figure}

After confirming that both contacts behave identically by taking a full scan of the device, we will henceforth consider only one of the metal contacts. Fig.\ref{graph:fig2} shows the influence of the excitation wavelength on photovoltage for a pn-junction configuration (p-doping of graphene underneath the contact and n-doping in the graphene channel). The photovoltage maps at short and long wavelengths are very different. For 785nm excitation (Fig.\ref{graph:fig2}b) the photovoltage is mostly generated close to the contact edge where the pn-junction is located, and vanishes where the graphene flake ends underneath the contact, indicated by the dotted line. At the corners of the metal contact a hot spot of enhanced responsivity occurs, due to curvature induced electric-field enhancement (lightning-rod effect)\cite{Ermushev1993}. On the other hand, at 457nm (Fig.\ref{graph:fig2}a) the whole contact area contributes to the photovoltage, with maxima at the contact edges. Even far away from the pn-junction located at the edge of the metal contact, a photovoltage is produced. This persists in the metal contact even when graphene is absent underneath, as such extending beyond the indicated dotted line.

\begin{figure}
\centerline{\includegraphics[width=90mm]{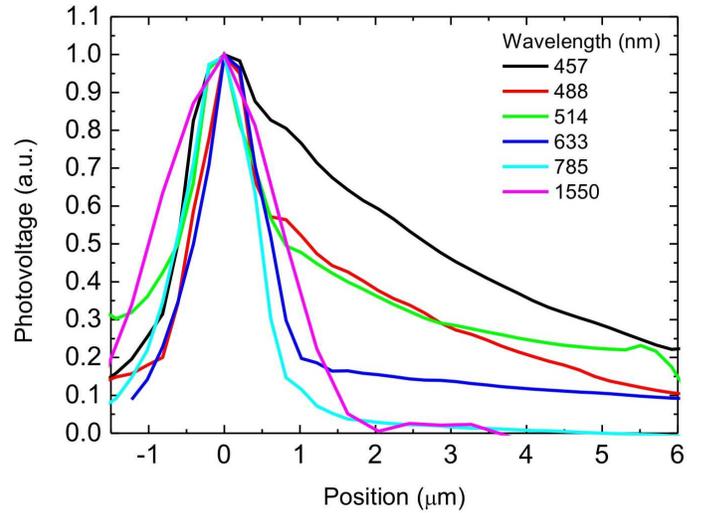}}
\caption{Spatial profile of photovoltage in the center of the metal contact.}
\label{graph:fig2c}
\end{figure}

Fig.\ref{graph:fig2c} shows that at 633, 785, and 1550nm a single peak at the contact edge is observed. Wavelengths of 457, 488, 514 nm lead to an additional decay of the photovoltage into the metal contact, with increasing decay lengths for shorter wavelengths.
\begin{figure}
\centerline{\includegraphics[width=90mm]{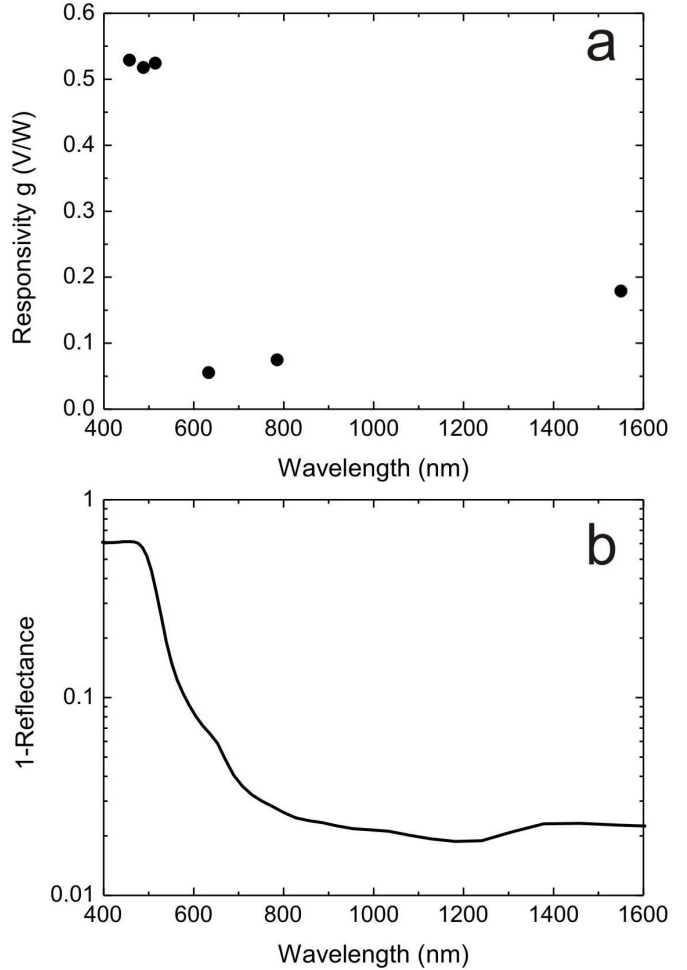}}
\caption{a) Responsivity of MGM-photodetector and b) 1-Reflectance of Au, as a function of excitation wavelength}
\label{graph:fig3}
\end{figure}

We now consider the dependence of the responsivity, g, on excitation wavelength, Fig.\ref{graph:fig3}a. This shows an enhancement towards shorter wavelengths (2.5 times bigger at 457nm compared to 1550nm). We assign this to an increased absorption of the incident light at shorter wavelengths on the Au contact. This leads to a T rise on the metal contact, thus heating the pn-junction at the contact edge, producing a thermoelectric contribution to the photovoltage. This is consistent with the photovoltage contribution of the metal contact far away from the contact edge, as seen in Fig.\ref{graph:fig2}a, because no light is incident directly on the junction. Considering Au's good ($\sim300\frac{W}{mK}$)\cite{Tritt2004} and graphene's excellent (up to$\sim5000\frac{W}{mK}$)\cite{Balandin2008} thermal conductivities, we assume that heat is transported to the pn-junction from within the metal contact, leading to a T gradient across the device and producing a thermoelectric contribution to the photovoltage\cite{Zuev2009,Wei2009}. Indeed, the trend in Fig.\ref{graph:fig3}a follows that of the heat energy $Q$ [J] deposited into the metal by the incident laser. This can be expressed as $Q\propto P_{abs}=(1-R) P_{input}$\cite{Chrisey1994}, with $P_{abs}[W]$ the absorbed power in the metal film, $R$ the Au film reflectance and $P_{input}$, the incident laser power. Fig.\ref{graph:fig3}b plots the calculated dependence of 1-R as a function of wavelength. For metals, the normal incidence R can be written as\cite{Kaye1995}:
\begin{equation}
R = \frac{(n_1-n_0)^2 + k_1^2}{(n_1+n_0)^2 + k_1^2}
\label{R}	
\end{equation}
where $n_0$ is the refractive index of the entrance medium, and $n_1$ and $k_1$ are the real and imaginary parts of the complex index of refraction of the absorbing medium. Taking Au's complex index of refraction from Ref.\cite{Palik1985} and considering that the entrance medium is air with $n_0 = 1$, the factor 1-R, proportional to the absorbed heat energy, is in good agreement with Fig\ref{graph:fig3}a. 1-R increases a factor 20 from 1550 to 457nm and explains the enhanced responsivity at shorter wavelengths as due to pronounced thermoelectric effects resulting from the Au contact heating. Even in the absence of graphene underneath the contact, a photoresponse is generated as the Au film spreads the heat energy towards graphene.
\begin{figure}
\centerline{\includegraphics[width=90mm]{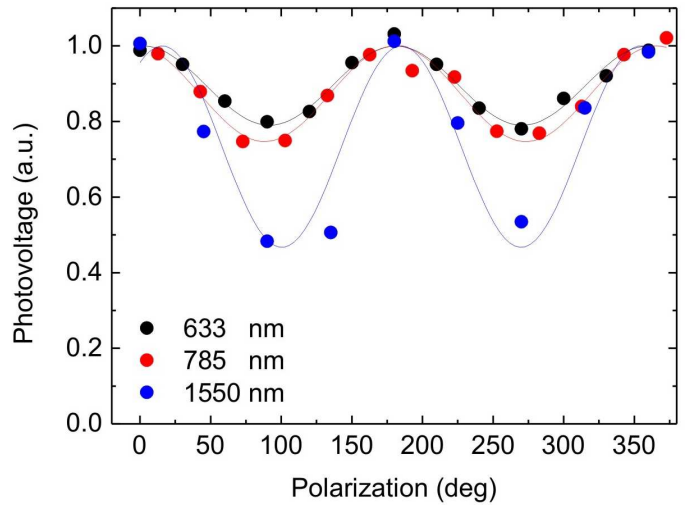}}
\caption{Dependence of photovoltage on incident polarization}
\label{graph:fig4a}
\end{figure}

To further investigate the influence of thermo- and photoelectric effects on the overall photovoltage, we perform polarization dependent measurements. Photovoltage maps are acquired at different polarization angles of the incident light, for a given location at the contact edge. Fig.\ref{graph:fig4a} plots the photovoltage in dependence of polarization at 633, 785, and 1550nm excitations. An angle of 0$^{\circ}$ denotes a polarization perpendicular to the metal contact edge. This shows two contributions: one polarization dependent, and another polarization independent.
\begin{figure}
\centerline{\includegraphics[width=80mm]{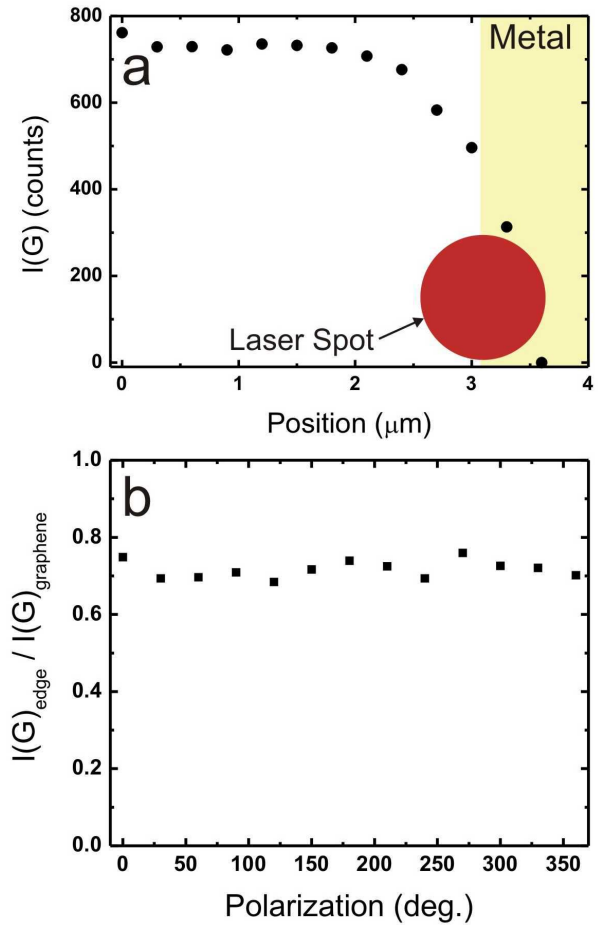}}
\caption{a) Linescan of I(G) approaching the contact edge. b) Polarization dependence of I(G) at the edge relative to I(G) away from the edge}
\label{graph:fig6}
\end{figure}

This behavior could be in principle due to plasmonic effects\cite{Echtermeyer2011,Schedin2010}. Polarization dependent excitation of plasmons at the metal edge could lead to near-field enhancement, thus a polarization dependent responsivity enhancement. Raman spectroscopy is used to investigate the influence of the metal electrode on a possible plasmonic dependence on light polarization. Spectra are first taken approaching the metal electrode from the bare graphene in a line scan with 300nm steps. Fig.\ref{graph:fig6}a plots I(G) as a function of position, for polarization perpendicular to the metal edge (maximum photovoltage). I(G) decreases as the metal edge is approached and no enhancement in the vicinity of the edge is observed. Instead, the metal electrode shields the light, resulting in a I(G) reduction. Polarization dependent Raman measurements are then carried out at the metal edge, Fig.\ref{graph:fig6}b. No trend is observed for the angular dependence. Another possible explanation could be surface plasmon polaritons (SPPs) that propagate from within the metal contact towards the junction at the edge of the contact\cite{Echtermeyer_SPP_2014}. However, experiments in combination with theoretical calculations demonstrate that SPPs cannot be excited on a flat metal contact\cite{Echtermeyer_SPP_2014}. Thus, plasmonic effects cannot explain the observed photovoltage angular dependence.
\begin{figure}
\centerline{\includegraphics[width=90mm]{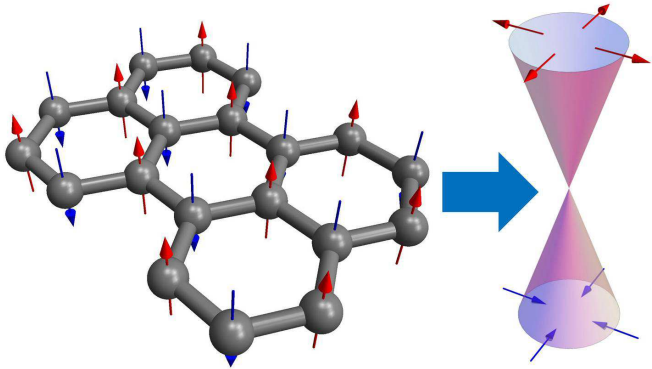}}
\caption{(Left) Honeycomb lattice of graphene and corresponding real space pseudospin orientation of the two interpenetrating Bravais sublattices, denoted in red and blue, respectively, and (Right) translation to momentum space.}
\label{graph:fig6h}
\end{figure}

We thus assign the polarization dependent contribution to the photoelectric effect, as due to polarization dependent interband optical excitations. Charge carriers in graphene are the $\pi$-electrons moving on a honeycomb lattice composed of {\em two} interpenetrating hexagonal sublattices. The sublattice degree of freedom is commonly associated with the pseudospin\cite{McCann2012}, which relates to the relative amplitude of the electron wave function located on either sublattice. If all electrons were placed on the ``red'' sublattice in Fig.\ref{graph:fig6h}, the pseudospin would be pointing upwards out of the SLG, whereas electrons on the ``blue'' sublattice correspond to pseudospin pointing downwards out of the SLG. Since both  ``red'' and ``blue'' lattice sites are occupied by the same carbon atoms, the electron density is distributed equally between these sublattices. Thus, the in-plane pseudospin orientation is determined by the phase difference in the amplitudes on the ``red'' and ``blue'' sites\cite{McCann2012}. The resulting texture is radial, as shown in Fig.\ref{graph:fig6h}.

The pseudospin-locked carriers can be described by the effective Hamiltonian\cite{McCann2012}: $H_0=v_F \vec{\sigma}\cdot \mathbf{p}$, where $\mathbf{p}$ is the two-component momentum, $\vec{\sigma}$ is the pseudospin, and $v_F\approx 10^6 \mathrm{ms^{-1}}$ is the Fermi velocity. The pseudospin texture represents the expectation value of the pseudospin operator $\vec{\sigma}$ with respect to the eigenstates of $H_0$\cite{McCann2012}.

\begin{figure}
\centerline{\includegraphics[width=85mm]{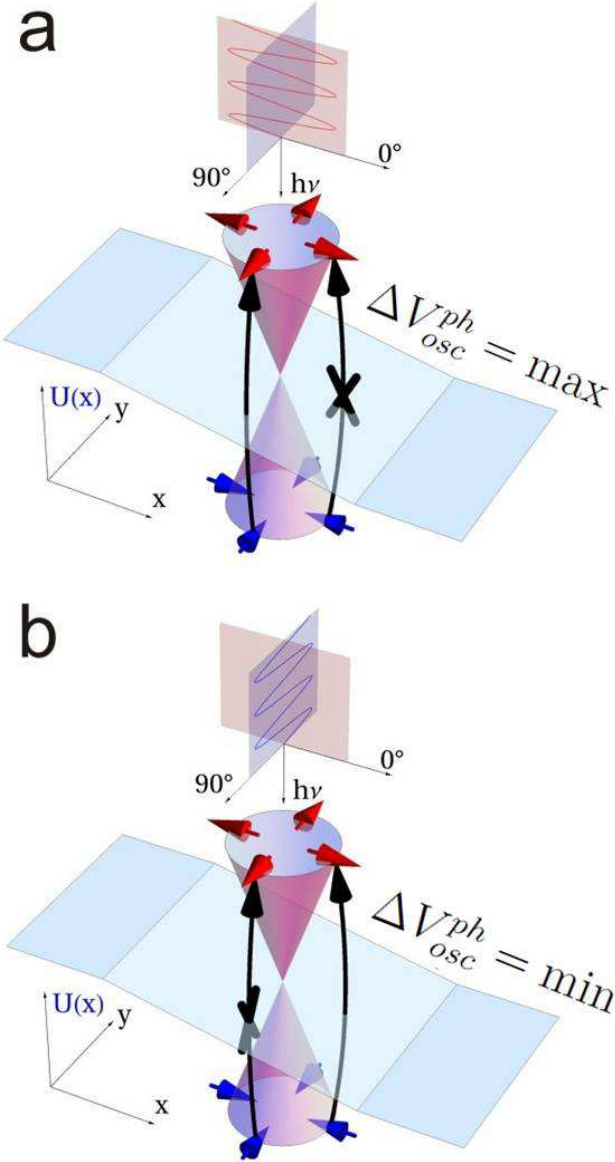}}
\caption{a,b) Linearly polarized light incident on a potential step $U(x)$ in graphene and pseudospin-dependent selection rule for interband optical excitations for a) $\theta_{pol}=0^\circ$ and b) $\theta_{pol}=90^\circ$. The polarization is characterized by the electric field ${\mathbf E}$. The photocarrier generation rate is anisotropic and proportional to $\sin^2(\widehat{{\mathbf E},{\mathbf p}})$, where ${\mathbf p}$ is the electron momentum. The driving term of the Boltzmann equation acting on a function of ${\mathbf p}/p$ is maximal when the force $-\nabla U(x)$ is perpendicular to the direction of motion, maximizing the photoresponse at $\theta_{pol} = 0^\circ$,
see the main text.}
\label{graph:fig4b}
\end{figure}

To excite an electron from the valence to the conduction band it is necessary to flip the pseudospin, as it is seen from Fig.\ref{graph:fig6h}. The interaction Hamiltonian between the charge carriers in graphene and an electromagnetic wave is characterized by the electric ${\mathbf E}=-\frac{1}{c}\frac{\partial \mathbf{A}}{\partial t}$ and magnetic ${\mathbf B}=\nabla\times\mathbf{A}$ fields, with $\mathbf{A}$ the vector potential. This can be derived form $H_0$ by substituting $\mathbf{p} \to \mathbf{p} -\frac{e}{c}\mathbf{A}$: $H_\mathrm{int}= \frac{e v_F}{c}\vec{\sigma}\cdot \mathbf{A}$. Assuming a linearly polarized electromagnetic wave with $\mathbf{A}=\mathbf{A}_0 \cos(\omega t- k z)$, the corresponding electric field is ${\mathbf E}=\mathbf{E}_0\sin(\omega t- k z)$, with $\mathbf{E}_0=\frac{\omega \mathbf{A}_0}{c}$, $\omega=2\pi c/\lambda$ the radiation frequency, and $k$ the normal component of the wave vector. Considering the commutator $[H_\mathrm{int},\vec{\sigma}]$, we get that these two operators commute with each other if and only if ${\mathbf A}$ (or ${\mathbf E}$) is along $\vec{\sigma}$. The pseudospin is then conserved and interband transitions are forbidden, as for Figs.\ref{graph:fig4b}a,b. In contrast, the commutator $[H_\mathrm{int},\vec{\sigma}]$ is maximum for ${\mathbf E} \perp \vec{\sigma}$, resulting in an interband transition rate maximum, Figs.\ref{graph:fig4b}a,b. Note that $\vec{\sigma}\parallel \mathbf{p}$ because of the pseudospin-momentum locking, Figs.\ref{graph:fig4b}a,b. As consequence, the photovoltage $V^\mathrm{ph}$ measured on the irradiated junction depends on the relative orientation between the polarization plane of the incident light and the junction.

The relaxation of photoexcited carriers to equilibrium in graphene consists of three processes with three characteristic time scales\cite{Sun2010,Natcomm2013brida,George2008,APL2008dawlaty,APL2012malic,PRB2011malic,PRB2013Tomadin}: In the first step, photoexcited carriers lose energy through e-e scattering on a$\sim$10fs time-scale\cite{Natcomm2013brida,PRB2013Tomadin}. Subsequently, this distribution thermalizes through electron-phonon (e-ph) scattering towards a hot Fermi-Dirac distribution\cite{Sun2010,Natcomm2013brida,George2008,APL2008dawlaty,APL2012malic,PRB2011malic,Song2012PRL,Graham2012,PRB2013Tomadin}, with time-scale in the range of hundreds of fs ($\tau_1$)\cite{Sun2010,Natcomm2013brida,George2008,APL2008dawlaty,APL2012malic,PRB2011malic,Song2012PRL,Graham2012,PRB2013Tomadin}. Finally, the hot Fermi-Dirac distribution relaxes to equilibrium by e-h recombination, which can lead to plasmon emission, phonon emission and Auger scattering on a ps timescale ($\tau_2$)\cite{Sun2010,Natcomm2013brida,George2008,APL2008dawlaty,APL2012malic,PRB2011malic,PRB2013Tomadin}.

The optical fluence $\Phi$ applied to our devices is low compared to that used in most pump-probe measurements, such as those in Refs.\cite{PBB2011breusing,Natcomm2013brida,George2008,APL2008dawlaty}. It is instructive to translate the fluence to the photoelectron concentration $n^\mathrm{ph}=\frac{\pi e^2}{\hbar c}\frac{\Phi}{\hbar\omega}$ with $\frac{\pi e^2}{\hbar c}=0.023$ the SLG optical absorption\cite{Nair2008}, or, equivalently,
$n^\mathrm{ph}=\frac{\pi e^2}{\hbar c}\frac{w_i \tau_2}{\hbar\omega}$, where $w_i=\frac{c|\mathbf{E}_0|^2}{8\pi}$ is the incident radiation power per square. The majority of pump-probe measurements were done for $n^\mathrm{ph}\sim10^{13}\,\mathrm{cm}^{-2}$\cite{PBB2011breusing,Natcomm2013brida,George2008,APL2008dawlaty}. In our case of continuous wave radiation with laser powers$\sim$few hundred $\mu \mathrm{W}$ and $\mu \mathrm{m}$ laser spot diameters, we have $n^\mathrm{ph}\sim10^{10}\,\mathrm{cm}^{-2}$, for a typical total recombination time $\tau_2\sim\mathrm{ps}$. The anisotropic distribution function for the photoexcited carriers relaxes to an hot Fermi-Dirac distribution at a T which could be much higher than room temperature, $T_\mathrm{room}$\cite{PBB2011breusing,Natcomm2013brida,APL2012malic,PRB2011malic}. In our case, however, the light induced photocarrier concentration is much lower than the always present intrinsic background electron concentration, even at zero chemical potential. The intrinsic e concentration $n$ at $T_\mathrm{room}$ can be calculated as $n=4\int\frac{d^2k}{4\pi^2}f^{(0)}_+(k,T_\mathrm{room})=\frac{\pi T_\mathrm{room}^2}{6\hbar^2v_F^2}$, with $f^{(0)}_+(k,T_\mathrm{room})$ the e-Fermi-Dirac distribution, and the multiplier $4$ due to the spin and valley degeneracy. This gives $n\sim10^{11}\,\mathrm{cm}^{-2}$, higher than $n^\mathrm{ph}\sim 10^{10}\,\mathrm{cm}^{-2}$ at the fluence used in our experiments. The same is true for the corresponding hole concentrations $p$ and $p^\mathrm{ph}$. The major contribution to the total carrier concentration thus originates from the intrinsic carriers, described by the Fermi-Dirac distribution at $T_\mathrm{room}$. In what follows, we therefore assume the characteristic e temperature to be $T_\mathrm{room}$.

Note that the role of carrier-carrier interactions in the relaxation of the photocarrier distribution is reduced substantially since the lower carrier concentration results in less frequent carrier-carrier collisions. As consequence, the ultrafast relaxation time scale$\sim10\,\mathrm{fs}$\cite{Natcomm2013brida} at high fluence is not considered here. We assume the relaxation of the anisotropic photocarrier distribution governed by scattering with optical phonons, rather than by e-e scattering, and characterized by the slower intraband relaxation rate $\tau_1$. This was measured$\sim150-170\, \mathrm{fs}$\cite{Natcomm2013brida}.

The interaction between the electromagnetic wave and charge carriers can be quantified using Fermi's golden-rule, considering $H_\mathrm{int}$ as a perturbation. The steady state distribution function is obtained by balancing the golden-rule e-h generation and relaxation rates. Since we are interested in the anisotropic part of the distribution function, the relevant relaxation time is $\tau_1=150\,\mathrm{fs}$ discussed above, rather than $\tau_2 > 1\,\mathrm{ps}$ associated with e-cooling and e-h recombination. The generation rate is proportional to $\sin^2(\theta-\theta_\mathrm{pol})$, where $\theta$ is the direction of e motion with $p_x=p\cos\theta$, $p_y=p\sin\theta$, and $\theta_\mathrm{pol}$ the polarization angle, see Methods.

The standard tool for photovoltage calculations is the drift-diffusion equation\cite{Nelson2004}, which considers e and h concentrations, rather than their distribution functions. The angular dependence of the distribution function is lost when the Boltzmann equation is integrated in momentum space to obtain the drift-diffusion relation\cite{Nelson2004}. To retain the angular dependence we have to take one step back and start from the Boltzmann equation:
\begin{equation}
\label{Boltzmann}
\mathbf{F}\cdot \nabla_\mathbf{p} f_\pm + \mathbf{v}\cdot \nabla_\mathbf{r} f_\pm = g^\mathrm{ph}_\pm -\frac{\Delta f_\pm}{\tau_1},
\end{equation}
where $\mathbf{F}$ is the electrostatic force, with $F_x=F\cos\phi$, $F_y=F\sin\phi$, $\mathbf{v}$ is the electron velocity, $g^\mathrm{ph}_\pm$ is the photogeneration rate, and $\Delta f_\pm = f_\pm - f^{(0)}_\pm$ is the deviation of the distribution function from the hot Fermi-Dirac distribution $f^{(0)}_\pm$ with ``$\pm$'' being the conduction/valence band index. The solution of this equation can be found in Methods. Having $f_\pm$ at hand we calculate the current density $j_x=4\sum_\pm\int \frac{dk^2}{4\pi^2} v_x f_\pm$, then set $j_x(V)$ to zero (open circuit) and extract the voltage $V$, which resembles the photovoltage $V^\mathrm{ph}$ in the absence of bias.

It is important to emphasize that the photovoltage maximum occurs for the perpendicular orientation of the light polarization plane with respect to the potential barrier when the majority of photoexcited electrons are moving {\em parallel} to the junction. This is due to the driving operator in the kinetic equation (Eq.\ref{Boltzmann}) which acts on the $\theta$-dependent steady state distribution function in a non-trivial way. The $\theta$-dependent part of the driving operator can be written as:
\begin{eqnarray}
 \nonumber \left(\mathbf{F}\cdot \nabla_\mathbf{p}\right)_\theta & = & \left( -F_x p_y +F_y p_x \right) \frac{1}{p^2}\frac{\partial}{\partial \theta} \\
& = &  \sin\left( \phi-\theta \right) \frac{F}{p}\frac{\partial}{\partial \theta},
\end{eqnarray}
where the relations $\frac{\partial \theta}{\partial p_x}=-p_y/p^2$ and $\frac{\partial\theta}{\partial p_y}=p_x/p^2$ have been utilized. Thus, the driving operator acting on the function of $\theta$ is maximum when the force and direction of particle motion are perpendicular, i. e. $\phi-\theta =\pi/2$. Thus, the major contribution to $\Delta f_\pm$ comes from e moving {\em parallel} to the barrier, photogenerated by the polarized light with $\theta_\mathrm{pol}=0^\circ$, as shown in Figs.\ref{graph:fig4b}a,b. The maximum photovoltage occurs therefore at $\theta_\mathrm{pol}=0^\circ$, not at $\theta_\mathrm{pol}=90^\circ$, as one might expect. A similar $90^\circ$ off-set was found in the photocurrent calculations of Ref.\cite{Mai2011}.

We distinguish two cases of n-n and p-n graphene junctions. The former is simpler and the resulting $\cos^2\theta_\mathrm{pol}$-dependent photovoltage term reads:
\begin{eqnarray}
\nonumber && -q V_\mathrm{osc}^\mathrm{ph}(\theta_\mathrm{pol}) =  \cos^2\theta_\mathrm{pol} \frac{\tau_1 \lambda^2 v_F^2}{2\pi c^2} \frac{W_a}{\pi d^2/4}\\
&& \times \ln \left(\frac{\mu_0- U\left(x+\frac{d}{2}\right)}{\mu_0-U\left(x-\frac{d}{2}\right)}\right).
\label{eq3}	
\end{eqnarray}
Here, $\lambda$ is the light wavelength, $\mu_0$ is the chemical potential in graphene in the absence of top metallic contacts, $U(x)$ is the built-in potential profile due to the metallic contacts, and the laser spot diameter is $d=1.5\,\mathrm{\mu m}$. To simplify the expression, we assume $\mu_0 - U(x)\gg T$ for any $x$, with $x$ the laser spot position. The absorbed radiation energy is characterized by the absorbed power $W_a=\frac{\pi e^2}{\hbar c} W_i$, which depends on the incident radiation power $W_i$, and SLG optical absorption $\frac{\pi e^2}{\hbar c}=0.023$. If the laser beam is focused on the middle of the n-n junction at $x=0$ and its size is larger than the junction region, then $\Delta U= U\left(\frac{d}{2}\right)-U\left(-\frac{d}{2}\right)$ is the built-in potential step forming the junction. The photovoltage depends weakly on $\Delta U$ and the logarithmic multiplier is smaller than $1$ for potential steps $\Delta U$ of a few 10s-100smeV, satisfying the $\mu_0 - U(x)\gg T$ criterion. This is different for the p-n junctions shown in Fig.\ref{graph:fig1pd}, where $\mu_0-U(x)\ll T$ in the middle of the junction.

\begin{figure}
\centerline{\includegraphics[width=90mm]{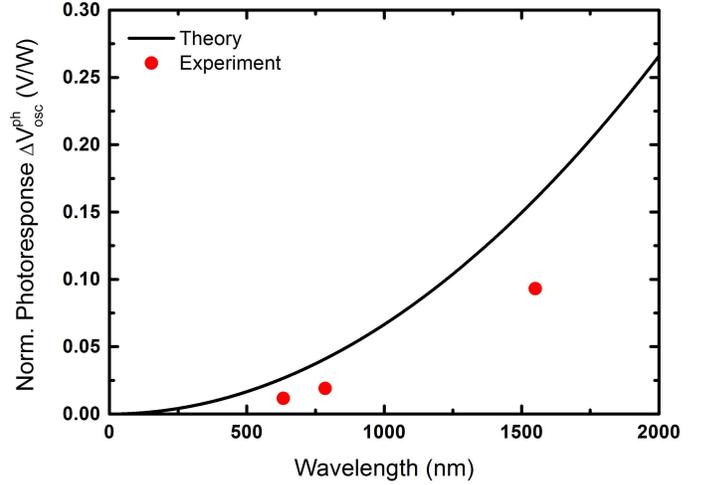}}
\caption{Experimental and theoretical polarization dependent photovoltage amplitude $\Delta V_\mathrm{osc}^\mathrm{ph}$ for our graphene p-n junction as a function of incident light wavelength. To fit the data, Eq.~(\ref{final}) is used with the following parameters: $\tau_1=150\,\mathrm{fs}$, $v_F = 10^6 \mathrm{ms^{-1}}$, $d=1.5\,\mathrm{\mu m}$, $U\left(-\frac{d}{2}\right)=-100\,\mathrm{meV}$, $U\left(+\frac{d}{2}\right)=50\,\mathrm{meV}$, $T=25\,\mathrm{meV}$.}
\label{graph:fig5}
\end{figure}

In what follows we assume $\mu_0=0$ and the electrochemical potential characterized by $U(x)$ alone. Eq.~(\ref{eq3}) is then rewritten as:
\begin{eqnarray}
\nonumber &&
-q  V_\mathrm{osc}^\mathrm{ph}(\theta_\mathrm{pol}) =  \cos^2 \theta_\mathrm{pol}  \frac{\tau_1 \lambda^2 v_F^2}{2\pi c^2} \frac{W_a}{\pi d^2/4} \\&&
\times \int\limits_\frac{U\left(x-\frac{d}{2}\right)}{2T}^\frac{U\left(x+\frac{d}{2}\right)}{2T} \frac{d\xi}{\ln\left(2\mathrm{cosh}\xi \right)}.
\label{eq4}	
\end{eqnarray}
T appears in Eq.7 since the condition $U(x)\gg T$ utilized before cannot apply in the middle of the p-n junction, where $U=0$, Fig.\ref{graph:fig1pd}. At $x=0$ (i.e. laser spot in the middle of the junction) and $\frac{U(\pm d/2)}{2T}=1$ (i.e. a potential step of $100\,\mathrm{meV}$), the integral is$\sim$1. At the radiation power of $300\,\mu\mathrm{W}$ and $\lambda=633\,\mathrm{nm}$, the amplitude of $\cos^2 \theta_\mathrm{pol}$ oscillations is a few $\mathrm{\mu V}$. The photovoltage amplitude $\Delta V_\mathrm{osc}^\mathrm{ph}$ at $x=0$ is then given by:
\begin{equation}
 \label{final}
  \Delta V_\mathrm{osc}^\mathrm{ph}(\theta_\mathrm{pol}) =  \frac{\tau_1 \lambda^2 v_F^2}{2\pi|q|c^2} \frac{W_a}{\pi d^2/4}
\int\limits_\frac{U\left(-\frac{d}{2}\right)}{2T}^\frac{U\left(+\frac{d}{2}\right)}{2T} \frac{d\xi}{\ln\left(2\mathrm{cosh}\xi \right)}.
\end{equation}
This is our main theoretical outcome, and is computed for our device as a function of $\lambda$ in Fig.\ref{graph:fig5}.

Note that the photovoltage is higher for longer wavelengths. This is because light with longer wavelength, but same radiation power, can excite more e into the conduction band, resulting in a stronger photoresponse. $V^\mathrm{ph}$ also contains a polarization-independent term of the same order as $V_\mathrm{osc}^\mathrm{ph}$. It is however not possible to separate this term from the thermoelectric contributions, which are isotropic and independent of the incoming light polarization, due to the isotropy of graphene and the Au film. The oscillating, incident light polarization dependent part of Fig.\ref{graph:fig4a} is thus a direct proof of a photoelectric contribution to the overall generated photovoltage. The magnitude of these oscillations with respect to the overall photoresponse allows us to estimate the relative photoelectric contribution $P_\mathrm{pe}$ to the overall photoresponse. Fig.\ref{graph:fig4a} shows that $P_\mathrm{pe}$ is at least 21, 25, and 53 \% for 633, 785, 1550nm. The experimentally and theoretically determined amplitude of the photoelectric polarization dependent part $\Delta V_\mathrm{osc}^\mathrm{ph}$ is shown in Fig.\ref{graph:fig5}, which demonstrates an increase of photoelectric contributions towards longer wavelengths.

In conclusion, we demonstrated the influence of the orientation of the lateral pn-junction in graphene-based photodetectors with respect to the polarization of incident linearly polarized light. The angular dependence is in good agreement with theory and a proof that both photo-thermoelectric and photoelectric effects contribute to the photoresponse in MGM-photodetectors, with photoelectric effects becoming more pronounced at longer wavelengths. Further, we demonstrated that the light generated anisotropic distribution of carriers in momentum space can be observed in electrical measurements despite their relaxation on ultra-fast time scales ($\tau_1$). This might open the possibility for graphene-based photodetectors that can detect incident light and its polarization on ultra-fast time scales, overcoming the thus-far speed limiting time constant $\tau_2$.
\section{Acknowledgements}
We thank Andrey Klots for useful discussions. We acknowledge funding from EU Graphene Flagship (contract no.604391), ERC grants NANOPOTS, Hetero2D, a Royal Society Wolfson Research Merit Award, EU projects GENIUS, CARERAMM, EPSRC grants EP/K01711X/1, and EP/K017144/1, Nokia Research Centre Cambridge, Deutsche Forschungsgemeinschaft (DFG) GRK 1570.
\section{Methods}
\subsection{Kinetic equation for optically excited carriers in graphene with  a built-in potential $U(x)$}
The Boltzmann kinetic equation (\ref{Boltzmann}) introduced above for $\Delta f_\pm = f_\pm - f^{(0)}_\pm$ can be written as:
\begin{equation}
\label{Boltzmann2}
-q\frac{\partial V}{\partial x} \frac{\partial f_\pm}{\partial p_x} - \frac{\partial U}{\partial x} \frac{\partial f_\pm}{\partial p_x}
+v_\pm \frac{\partial f_\pm}{\partial x}= g^\mathrm{ph}_\pm -\frac{f_\pm - f^{(0)}_\pm}{\tau_1},
\end{equation}
where $p_x=\hbar k_x$, $v_\pm=\pm v_F k_x/k$, $q=-|q|$ is the electron charge, and $U(x)$ is the built-in potential. The bias voltage is assumed to be small enough to justify the linear response in terms of $\frac{\partial V}{\partial x}$. In the absence of bias and photogeneration, the system is in the equilibrium state described by the Fermi-Dirac distribution function $f^{(0)}_\pm$:
\begin{equation}
\label{solution1}
f^{(0)}_\pm=\frac{1}{1+\exp\left(\frac{\pm\hbar v_F k + U(x) - \mu_0}{T} \right)}.
\end{equation}
``$\pm$'' stands for the conduction and valence band. The photogeneration rate $g^\mathrm{ph}_\pm$ can be derived from Fermi's golden rule using the unperturbed eigenstates of $H_0$, $\psi_{\pm\mathbf{k}}(x,y)=\frac{1}{\sqrt{2}}{\mathrm e}^{ik_x x+ik_y y} \left(1,\pm {\mathrm e}^{i\theta}\right)^T$, where $\tan\theta=k_y/k_x$. For a given spin/valley channel we get:
\begin{eqnarray}
 \nonumber && g^\mathrm{ph}_\pm = \frac{2\pi}{\hbar}\left(\frac{\hbar q v_F E_0}{2\hbar\omega}\right)^2\sin^2\left(\theta_\mathrm{pol}-\theta\right)\\
 && \times \delta(\hbar\omega-2\hbar v_F k)(f^{(0)}_\mp - f^{(0)}_\pm),
\end{eqnarray}
where $\theta_\mathrm{pol}$ is the polarization angle, and $E_0$ is the electric field amplitude of the electromagnetic wave, which can be related to the incident radiation power per unit square as $w_i=\frac{c}{4\pi}\langle [\mathbf{E}\times\mathbf{B}]_z \rangle_t=\frac{c E_0^2}{8 \pi}$ [W/cm$^2$]. The fluence can be estimated as $\Phi=w_i \Delta t$. The integral radiation power is $W_i=w_i\frac{\pi d^2}{4}$, where $d$ is the laser spot diameter.

We look for the solution of Eq.\ref{Boltzmann2} in the form $f_\pm=f^{(0)}_\pm+f^\mathrm{ph}_\pm+f^{(1)}_\pm$, where $f^\mathrm{ph}_\pm=\tau_1 g^\mathrm{ph}_\pm$, and $f^{(1)}_\pm$ is determined from the following equation obtained by the substitution of $f_\pm$ in Eq.\ref{Boltzmann2}:
\begin{eqnarray}
 \label{Boltzmann3}
 &&
 -q\frac{\partial V}{\partial x} \frac{\partial f^{(0)}_\pm}{\partial p_x} - \frac{\partial U}{\partial x} \frac{\partial f^{(0)}_\pm}{\partial p_x}
+v_\pm \frac{\partial f^{(0)}_\pm}{\partial x} \\
 \nonumber && -q\frac{\partial V}{\partial x} \frac{\partial f^\mathrm{ph}_\pm}{\partial p_x} - \frac{\partial U}{\partial x} \frac{\partial f^\mathrm{ph}_\pm}{\partial p_x}
+v_\pm \frac{\partial f^\mathrm{ph}_\pm}{\partial x}
= -\frac{f^{(1)}_\pm}{\tau_1}.
\end{eqnarray}
But, $- \frac{\partial U}{\partial x} \frac{\partial f^{(0)}_\pm}{\partial p_x} +v_\pm \frac{\partial f^{(0)}_\pm}{\partial x}=0$.
Moreover, $-q\frac{\partial V}{\partial x} \frac{\partial f^\mathrm{ph}_\pm}{\partial p_x}\ll-q\frac{\partial V}{\partial x} \frac{\partial f^{(0)}_\pm}{\partial p_x}$,
since $n^\mathrm{ph}\ll n$, as discussed in the main text. Taking into account the $\theta$ dependence of $f^\mathrm{ph}_\pm$, $f^{(1)}_\pm$ can be written as:
\begin{eqnarray}
\nonumber && \frac{f^{(1)}_\pm}{\tau_1}=q\frac{\partial V}{\partial x}\frac{\partial f^{(0)}_\pm}{\partial p_x}
 + \frac{\partial U}{\partial x}\frac{\pi\tau_1 v_F}{2\hbar}\left(\frac{ q  E_0}{2\hbar\omega}\right)^2(f^{(0)}_\mp - f^{(0)}_\pm)\\
\nonumber && \times  \left\{ \cos\theta \left[1-\cos\left(2\theta_\mathrm{pol}-2\theta\right)\right]\left[\frac{\partial}{\partial k}\delta\left(k-\frac{\omega}{2v_F}\right)\right] \right. \\
 && \left. + 2\sin\theta\sin\left(2\theta_\mathrm{pol}-2\theta\right)\frac{1}{k}\delta\left(k-\frac{\omega}{2v_F}\right)
 \right\}.
 \label{solution3}
\end{eqnarray}
Note the graphene specific contribution proportional to $\frac{\partial \sin^2(\theta-\theta_\mathrm{pol})}{\partial k_x}=\sin(2\theta_\mathrm{pol}-2\theta)\frac{\sin\theta}{k}$. To calculate the current density we multiply Eq.~(\ref{solution3}) by $v_\pm$ and integrate it over $k$ and $\theta$. We assume that $f^{(0)}_+(k=\frac{\omega}{2v_F}) = 0$ and $f^{(0)}_-(k=\frac{\omega}{2v_F}) = 1$, as reasonable for any T, electrochemical doping and wavelength we consider in this paper. In order to find the photovoltage for the open circuit we employ in our measurements, the total current density and external bias are set to zero. In this case, $V$ in Eq.~(\ref{solution3}) is the photovoltage $V^\mathrm{ph}$.
\subsection{Photoresponse of graphene n-n junction}
Here we assume that $\mu_0-U(x)\gg T$, so that Eq.\ref{solution3} can be integrated:
\begin{equation}
\label{nn}
q\frac{\partial V^\mathrm{ph}}{\partial x} (\mu_0-U(x)) +(2+\cos2\theta_\mathrm{pol}) \frac{\partial U}{\partial x}
\frac{\pi\tau_1}{2\hbar}\left(\frac{q v_F E_0}{2\omega}\right)^2 = 0.
\end{equation}
To obtain $V^\mathrm{ph}$ we integrate Eq.\ref{nn} over $x$ within the laser spot:
\begin{eqnarray}
&& \nonumber -q V^\mathrm{ph} = (2+ \cos2\theta_\mathrm{pol}) \frac{\pi\tau_1}{2\hbar}\left(\frac{q v_F E_0}{2 \omega}\right)^2 \\
&&\label{vph1} \times \int\limits_{x-\frac{d}{2}}^{x+\frac{d}{2}} dx \frac{1}{\mu_0-U(x)} \frac{\partial U}{\partial x}.
\label{DelVph}
\end{eqnarray}
The $E_0$-dependent multiplier in Eq.\ref{DelVph} can be expressed in terms of the absorbed radiation power $W_a=\frac{\pi e^2}{\hbar c} W_i$ (or absorbed radiation power per square  $w_a=\frac{\pi e^2}{\hbar c} w_i$). Extracting the $\theta_\mathrm{pol}$ dependent part out of Eq.\ref{DelVph}, we arrive at the final result of Eq.\ref{eq3}.
\subsection{Photoresponse of graphene p-n junction}
Here the electrochemical potential can be smaller than the junction region T, and we cannot assume $\mu_0-U(x)\gg T$. To simplify, we set $\mu_0=0$, so that the electrochemical potential is determined by $U(x)$ alone. Note that Eq.\ref{nn} is now $T$ dependent:
\begin{eqnarray}
&& \nonumber
q \frac{\partial V^\mathrm{ph}}{\partial x} 2T\ln\left(2\mathrm{cosh}\frac{U(x)}{2T} \right) \\
&&\nonumber
 + (2+\cos2\theta_\mathrm{pol}) \frac{\partial U}{\partial x}
\frac{\pi\tau_1}{2\hbar}\left(\frac{ q v_F E_0}{2\omega}\right)^2 = 0.\\
\label{pn}
\end{eqnarray}
and the photovoltage becomes:
\begin{eqnarray}
&& \nonumber
-q V^\mathrm{ph} = (2+ \cos2\theta_\mathrm{pol}) \frac{\pi\tau_1}{2\hbar}\left(\frac{q v_F E_0}{2\omega}\right)^2 \\
&& \times
\int\limits_{x-d/2}^{x+d/2} \frac{dx}{2T} \frac{1}{\ln\left(2\mathrm{cosh}\frac{U(x)}{2T} \right)} \frac{\partial U}{\partial x}.
\label{DelVph2}
\end{eqnarray}
One can exclude $T$ from the integrand. The final formula for $V^\mathrm{ph}$ reads:
\begin{eqnarray}
 \label{vph5}
\nonumber && -e V^\mathrm{ph} =(2+ \cos2\theta_\mathrm{pol}) \frac{\pi\tau_1}{2\hbar}\left(\frac{ e v_F E_0}{2\omega}\right)^2\\
&&
\times\int\limits_{-\frac{U(x-d/2)}{2T}}^\frac{U(x+d/2)}{2T} \frac{d\xi}{\ln\left(2\mathrm{cosh}\xi \right)}.
\end{eqnarray}
Extracting the $\theta_\mathrm{pol}$ dependent part from Eq.\ref{vph5}, we get Eq.\ref{eq4}.
\subsection{Thermoelectric contribution in the total photoresponse}
An irradiated sample experiences heating, therefore the electrons are subject to a T gradient $\frac{\partial T}{\partial x}$, which appears in Eq.\ref{Boltzmann2}, when $v_\pm \frac{\partial f_\pm}{\partial x}$ is written explicitly. Following the same procedure as above, we arrive at Eq.\ref{Boltzmann3}, where
$- \frac{\partial U}{\partial x} \frac{\partial f^{(0)}_\pm}{\partial p_x}+v_\pm \frac{\partial f^{(0)}_\pm}{\partial x}$ is not zero, and gives the leading contribution in terms of $\frac{\partial T}{\partial x}$. This cannot depend on light polarization in any circumstance.

For Eq.\ref{nn}, the thermoelectric term can be estimated as $\frac{\pi^2}{3} T(x)  \frac{\partial T}{\partial x}$, which results in the photothermoelectric term given by Eq.\ref{eq1}. Thus, the thermoelectric contribution, being proportional T, gets larger for hot electrons, and becomes dominant in this case. Most importantly, the thermoelectric response mainly depends on the radiation power converted to heat, and is not sensitive to any particular light polarization. In contrast, the photoelectric response (Eq.\ref{DelVph}) {\em does} depend on the polarization angle $\theta_\mathrm{pol}$, which makes it possible to separate these two effects in the total response measured. Note, however, that the photovoltage (Eq.\ref{DelVph}) also contains a $\theta_\mathrm{pol}$ independent contribution, which is not possible to distinguish from the thermoelectric response. Nevertheless, the amplitude of $\cos 2\theta_\mathrm{pol}$ oscillations gives indication of how large the photoelectric response is.

\end{document}